# Raman Beam Cleanup in Silicon in the Mid-Infrared

David Borlaug[1], Robert Rice[2], and Bahram Jalali[1]

[1]*Electrical Engineering Department, University of California, Los Angeles, California 90025, USA*
[2]*Northrop Grumman Space Technology, One Space Park, Redondo Beach, CA 90278, USA*

We report evidence of beam cleanup during stimulated Raman scattering in silicon. An amplified near-diffraction-limited Stokes beam is obtained from a severely aberrated pump beam.

Many high power lasers have poor beam quality resulting in excess beam divergence and low intensity when incident on a target. Beam cleanup can be used to convert a high power, low beam quality source to a high power, high beam quality source with higher far-field intensity[1-9]. A result of nonlinear optical processes, beam cleanup is a phenomenon in which a spatially aberrated pump beam transfers power to a spatially clean signal beam. As long as the coupling of energy to the clean signal beam is more favorable than to amplified spontaneous emission, the signal beam will have better beam quality than the pump, as characterized by a lower $M^2$ number ($M^2$ is a widely accepted metric for describing beam propagation and far-field characteristics[2]). Beam cleanup has been successfully observed in multimode fibers,[1] bulk gain samples[2-8] using two-wave mixing,[5-6] using four-wave mixing,[8] using fiber-based Raman scattering,[2-4] and fiber-based Brillouin scattering.[1] The behavior of Raman beam cleanup as a function of gain, pump, and Stokes powers has also been numerically studied.[9]

The experimental setup used to measure beam cleanup in silicon using stimulated Raman scattering (SRS) is shown in figure 1.

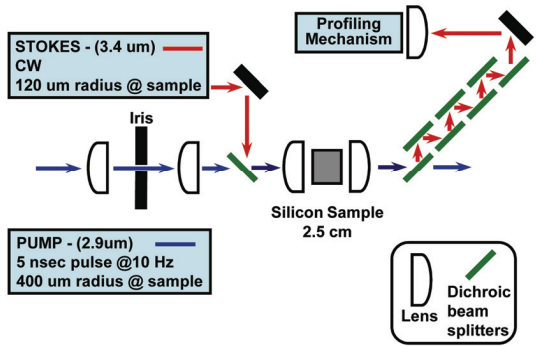

FIG. 1. (Color online) Experimental setup.

The Stokes signal was a continuous-wave HeNe laser with a wavelength of 3.4 μm and 2 mW output power. The pump was a Nd:YAG pumped tunable OPO with a Raman shifted wavelength of 2.9 μm, 5 ns FWHM pulses, and a 10 Hz repetition rate. The pump energy was 0.9 mJ resulting in an intensity of approximately 80 MW/cm$^2$ where the pump and Stokes beams overlapped. An iris was used to eliminate the possibility of pump spatial translation before the silicon sample. A dichroic beam combiner was used to combine the pump and Stokes beams which are then focused onto an anti-reflection coated 2.5 cm bulk silicon sample (5 cm diameter cylinder with a depth of 2.5 cm). The output radiation was collimated with a lens immediately after the sample.

The beams were characterized using a custom beam profiling mechanism consisting of an interchangeable pinhole analyzer, an InAs photodetector (Judson J12-18C-R01M), and computer controlled translation stages with 5 μm spatial accuracy (Newport LTA-HS). The computer was used to record data, move the stages, and to trigger an oscilloscope at the pump repetition rate. The pinhole size can be changed to profile high- and low-power beams without causing detector saturation. The detector had a 15 ns response time and 256 pulses were averaged at each measurement position. The following formula [Eq. (1)] can be easily derived from Ref. 10 and is used here for extracting the $M^2$ parameter from the measured profiles. For brevity and due to laboratory restraints (time and samples) all measurements produced $M_x^2$ values and are written succinctly as $M^2$. Two line scans are used to compute $M^2$ using equation 1: the line scan at the waist of the beam, and the line scan furthest from the waist where $z_r = \pi w_0^2/\lambda$.

$$M_x^2 = \frac{4\pi \sigma_x(z=0)\sigma_x(z \gg z_r)}{z\lambda}$$

$$\sigma_x^2 = \frac{\int_{-\infty}^{\infty}(x-x_0)^2 I(x,y=0,z)dx}{\int_{-\infty}^{\infty} I(x,y=0,z)dx} \quad (1)$$

Measuring in the X-Z plane ($M_x^2$) drastically reduces the measurement space, critical for a 10 Hz repetition rate laser. Furthermore, pinhole line scans are similar to the common "knife-edge" scans used for $M^2$ measurements. Careful beam alignment ensured that no beam walkoff occurred over the measurement space.

The measured profile of the pump beam at the silicon sample is shown in figure 2. The data yields $M^2$=28.6, a high value M² value indicating a very non-Guassian beam, expected for the particular type of pump laser used.

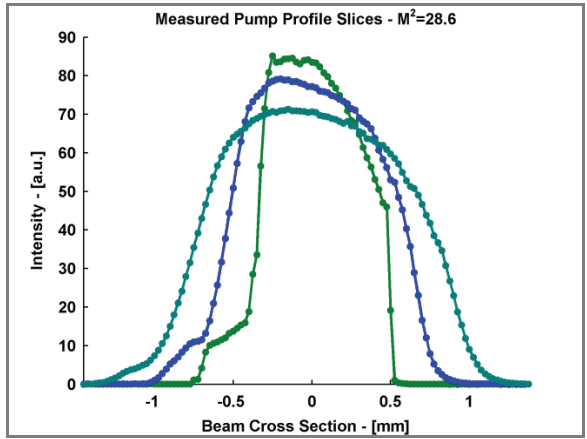

FIG. 2. (Color online) Beam cross sections measured as the beam propigates in the axial z-direction. The data yields $M^2=28.6$ for the pump beam.

The amplified and unamplified Stokes profiles and $M^2$ values were similarly obtained using the profiling mechanism. In order to spectrally filter the Stokes from the pump a non-distortive filter is needed. A grating spectrometer can not be used as it distorts the optical beam during diffraction and clips the beam at the entrance and exit slits of the spectrometer. Therefore, multi-stage dichroic filters were used to filter the pump and to obtain sufficient rejection. The power reflected by each filter was 98% at the Stokes wavelength and 4.5% at the pump wavelength, offering a rejection ratio of $4\times10^{-10}$ for a cascade of seven filters. Dichroic filters do not suffer from beam clipping but like all filters are phase inference devices that cause aberration through distributed reflection. With the pump off, the unamplified Stokes beam quality was measured with and without the filters. It was observed that the addition of the filters causes the unamplified Stokes $M^2$ to increase from 1.7 to 3.3.

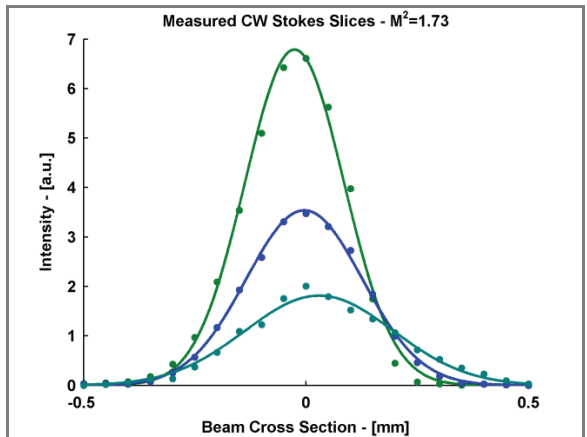

FIG. 3. (Color online) The unamplified Stokes $M^2$ parameter increases by 1.6 after seven dichroic filters. The unfiltered signal, shown above, is measured at the location of the silicon sample; the filtered beam is measured after the filters (see Fig. 1).

This increase of 1.6 in the $M^2$ parameter, caused by the filters and not by SRS amplification, must be taken into account when interpreting the beam quality of the amplified Stokes. The exact mathematical mapping of $M^2$ by dichroic filtering is unknown to the authors at this time. Therefore, as a simple approach to correct for the influence of the filters on the $M^2$ parameter, the contribution of the filters (i.e. a factor of 1.6) is subtracted from the measured $M^2$ value for filtered beams.

Once the influence of the filters on the $M^2$ value was understood, both the pump and Stokes were turned on and the amplified Stokes signal was profiled.

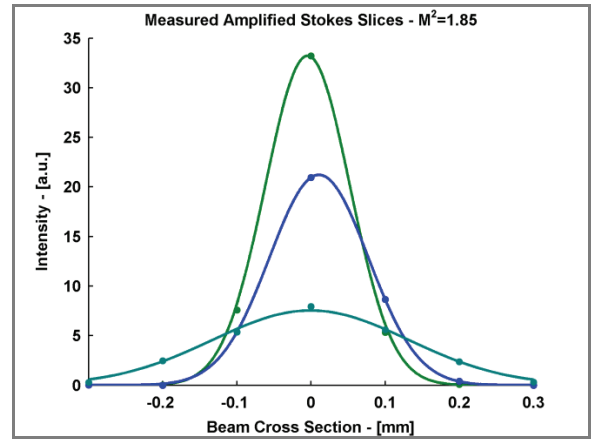

FIG. 4. (Color online) Amplified Stokes beam cross sections as the beam propigates in the axial z-direction. The data yields $M^2=3.4$. After accounting for an $M^2$ contribution of 1.6 by the dichroic filters, the true $M^2$ value for the amplified Stokes beam is estimated to be 1.8, compared to an $M^2$ of 28.6 for the pump.

The $M^2$ value of the amplified Stokes is 1.8, compared with an $M^2$ of 28.8 for the pump, demonstrating beam cleanup. The $M^2$ value of the amplified Stokes, 1.8, is nearly the same as that of unamplified Stokes, 1.7. The overall stokes on-off gain measured was 4.6 dB. Approximately 18,000 pump pulses were needed to acquire the complete beam profile. To avoid damaging the AR coating during exposure the experiment was conducted at a modest pump intensity of 80 MW/cm² resulting in low gain. Average Stokes gain as high as 12 dB has been measured using this laser at 3.1 mJ pump energy, vs. 0.9 mJ used in the present experiments.[11] However, such high pulse energies could not be used here due to AR coating damage during the long exposure.

Because of the low input Stokes power (2 mW) and the relatively modest gain (4.3 dB), the amount of pump-to-Stokes power transfer is negligible in the present experiments. The power transfer can be increased if a larger input Stokes power is used and when a larger Raman gain is realized. A question then arises as to whether the improvement in beam quality as evidenced by the drastic reduction in the pump vs. Stokes M$^2$ value observed here will continue to exist for larger values of gain and larger input Stokes powers. With respect to the input Stokes power, higher power is expected to improve the M$^2$ value even more as the relative influence of spontaneous emission in triggering SRS is expected to be less.[10] Higher pump power will increase the M$^2$ value, but the increase is expected to be modest for practically achievable gain values.[9]

This work was supported by a seedling grant from DARPA-MTO. The authors would like to thank Dr. Henryk Temkin for his support.